\documentclass{ws-ijmpa}

\begin{document}

\markboth{Isabella Amore et al.}
{NEMO: A Project for a km$^3$ Underwater Detector for Astrophysical Neutrinos}

%
\catchline{}{}{}{}{}
%

\title{NEMO: A PROJECT FOR A KM$^3$ UNDERWATER\\
DETECTOR FOR ASTROPHYSICAL NEUTRINOS\\
IN THE MEDITERRANEAN SEA\\}

\author{
I. Amore$^{4,5,}$\footnote{Corresponding author. Tel/Fax: +39 095 542 398. E-mail: amore@lns.infn.it (I. Amore).}   , S. Aiello$^{3}$, M. Ambriola$^{1}$, F. Ameli$^{11}$, M. Anghinolfi$^{7}$, A. Anzalone$^{5}$, G. Barbarino$^{8}$, E. Barbarito$^{1}$, M. Battaglieri$^{7}$, R. Bellotti$^{1}$, N. Beverini$^{10}$, M. Bonori$^{11}$, B. Bouhadef$^{10}$, M. Brescia$^{8,9}$, G. Cacopardo$^{5}$, F. Cafagna$^{1}$, A. Capone$^{11}$, L. Caponetto$^{3}$, E. Castorina$^{10}$, A. Ceres$^{1}$, T. Chiarusi$^{11}$, M. Circella$^{1}$, R. Cocimano$^{5}$, R. Coniglione$^{5}$, M. Cordelli$^{6}$, M. Costa$^{5}$, S. Cuneo$^{7}$, A. D'Amico$^{5}$, G. De Bonis$^{11}$, C. De Marzo$^{1,}$\footnote{Deceased}   , G. De Rosa$^{8}$, R. De Vita$^{7}$, C. Distefano$^{5}$,
E. Falchini$^{10}$, C. Fiorello$^{1}$, V. Flaminio$^{10}$, K. Fratini$^{7}$, A. Gabrielli$^{2}$, S. Galeotti$^{10}$, E. Gandolfi$^{2}$, G. Giacomelli$^{2}$, F. Giorgi$^{2}$, A. Grimaldi$^{3}$, R. Habel$^{6}$, E. Leonora$^{3,4}$, A. Lonardo$^{11}$, G. Longo$^{8}$, D. Lo Presti$^{3,4}$, F. Lucarelli$^{11}$, E. Maccioni$^{10}$, A. Margiotta$^{2}$, A. Martini$^{6}$, R. Masullo$^{11}$, R. Megna$^{1}$, E. Migneco$^{4,5}$, M. Mongelli$^{1}$, T. Montaruli$^{12,}$\footnote{On leave of absence from Dipartimento Interateneo di Fisica Universit\`a di Bari, Via E. Orabona 4, 70126, Bari, Italy.}   , M. Morganti$^{10}$, M.S. Musumeci$^{5}$, C.A. Nicolau$^{11}$, A. Orlando$^{5}$, M. Osipenko$^{7}$, G. Osteria$^{8}$, R. Papaleo$^{5}$, V. Pappalardo$^{5}$, C. Petta$^{3,4}$, P. Piattelli$^{5}$, G. Raia$^{5}$, N. Randazzo$^{3}$, S. Reito$^{3}$, G. Ricco$^{7}$, G. Riccobene$^{5}$, M. Ripani$^{7}$, A. Rovelli$^{5}$, M. Ruppi$^{1}$, G.V. Russo$^{3,4}$, S. Russo$^{8}$, P. Sapienza$^{5}$, M. Sedita$^{5}$, E. Shirokov$^{7}$, F. Simeone$^{11}$, V. Sipala$^{3,4}$, M. Spurio$^{2}$, M. Taiuti$^{7}$, G. Terreni$^{10}$, L. Trasatti$^{6}$, S. Urso$^{3}$, V. Valente$^{6}$, P. Vicini$^{11}$.\\}

\address{
$^1$INFN Sezione Bari and Dipartimento Interateneo di Fisica Universit$\grave{a}$ di Bari,  Via E. Orabona 4, 70126, Bari, Italy\\
$^2$INFN Sezione Bologna and Dipartimento di Fisica Universit$\grave{a}$ di Bologna, V.le Berti Pichat 6-2, 40127, Bologna, Italy\\
$^3$INFN Sezione Catania, Via S.Sofia 64, 95123, Catania, Italy\\
$^4$Dipartimento di Fisica and Astronomia Universit$\grave{a}$ di Catania, Via S.Sofia 64, 95123, Catania, Italy\\
$^5$INFN Laboratori Nazionali del Sud, Via S.Sofia 62, 95123, Catania, Italy\\
$^6$INFN Laboratori Nazionali di Frascati, Via Enrico Fermi 40, 00044, Frascati (RM), Italy\\
$^7$INFN Sezione Genova and Dipartimento di Fisica Universit$\grave{a}$ di Genova, Via Dodecaneso 33, 16146, Genova, Italy\\
$^8$INFN Sezione Napoli and Dipartimento di Scienze Fisiche Universit$\grave{a}$ di Napoli, Via Cintia, 80126, Napoli, Italy\\
$^9$INAF Osservatorio Astronomico di Capodimonte, Salita Moiariello 16, 80131, Napoli, Italy\\
$^{10}$INFN Sezione Pisa and Dipartimento di Fisica Universit$\grave{a}$ di Pisa, Polo Fibonacci, Largo Bruno Pontecorvo 3, 56127, Pisa, Italy\\
$^{11}$INFN Sezione Roma 1 and Dipartimento di Fisica Universit$\grave{a}$ di Roma "La Sapienza", P.le A. Moro 2, 00185, Roma, Italy\\
$^{12}$University of Wisconsin, Department of Physics, 53711, Madison, WI, USA}

\maketitle

\begin{history}
\received{Day Month Year}
\revised{Day Month Year}
\end{history}

\begin{abstract}
The status of the project is described: the activity on long
term characterization of water optical and oceanographic
parameters at the Capo Passero site candidate for the Mediterranean
km$^3$ neutrino telescope; the feasibility study; the physics performances and underwater technology for the km$^3$; the activity on NEMO Phase 1, a
technological demonstrator that has been deployed at 2000 m
depth 25 km offshore Catania; the realisation of an underwater infrastructure at 3500 m depth at the candidate site (NEMO Phase 2).

\end{abstract}

\ccode{PACS numbers: 29.40.Ka, 95.55.Vj, 95.85.Ry}

\section{Introduction}

The construction of a km$^3$ underwater detector for high energy astrophysical
neutrinos is one of the main goals of astroparticle physics.
High energy neutrinos are very promising probes for the investigation of the far universe and of the acceleration processes occurring in galactic and extragalactic sources. Differently from charged particles and gamma rays
with $E_\gamma > $TeV, neutrinos can reach the Earth from far away
cosmic accelerators, travelling in straight line, thus
carrying direct information on the source. Theoretical models
indicate that a detection area of $\simeq$1 km$^2$ is required for
the measurement of high energy neutrino cosmic fluxes. The underwater/ice
Cherenkov technique is considered the most promising
experimental approach to build high energy neutrino detectors.
The first generation of underwater/ice neutrino telescopes, BAIKAL\cite{BaikalICRC2003} and
AMANDA\cite{Ackermann05point}, despite their limited size have already set first
constraints on astrophysical models of TeV neutrino sources.
The successful experience of AMANDA opened the way to
IceCube\cite{ICECUBE2003}, a km$^3$-size neutrino telescope which
is now under construction at the South Pole. ANTARES\cite{ANTARES2006} has already deployed 8 lines, of which 5 are taking data since february 2007. A module of the NESTOR\cite{NESTOR} detector has been deployed in March 2003. The NEMO\cite{Migneco} collaboration
is carrying out, since 1998, an $R\&D$ programme for the construction of a km$^3$ detector and on December 2006 deployed a technological demonstrator (the NEMO Phase 1
project). The scientific
community is supporting the construction of another km$^3$
neutrino telescope in the Northern Hemisphere, to allow
contemporary observation of the full sky. The Mediterranean Sea
offers optimal conditions to locate the telescope and a
Mediterranean km$^3$ telescope should be able to observe neutrinos
from the Galactic Cente region, not seen by IceCube. For this reason the EU funded KM3NeT\cite{km3net} a design study activity for the Mediterranean km$^3$ telescope that involves several European scientific institutions which is expected to provide a technical design report within 2009. In this paper the activities of NEMO are described. They have been focused on
three items: the search and long term monitoring of an optimal
site for the installation of the km$^3$ telescope; the development
of a technical and scientific feasibility study of a km$^3$ detector;
the realisation the NEMO Phase 1 project.

\section{Site selection and characterization}

\begin{figure}[h]
\centerline{\psfig{file=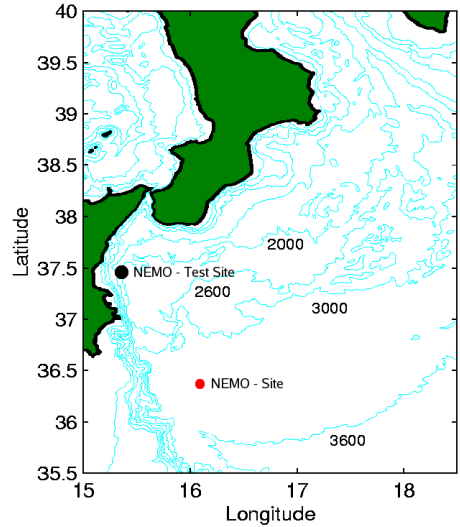,width=7 cm}}
\vspace*{8pt}
\caption{Bathymetric
chart of Eastern Sicily. The locations of the Capo Passero site and
of the NEMO Phase 1 Test Site are shown. The seabed depth is
about 3500 m for the Capo Passero site and 2000 m for the Test
Site. \label{fig:capo_passero}}
\end{figure}

The km$^3$ detector needs first a complete knowledge
of the site physical and oceanographical characteristics over a
long time period. The NEMO collaboration performed,
since 1998, a long term research program to find and
characterise an optimal deep sea site. After 30 sea campaigns, a
site located in the Ionian Sea (36$^\circ$ 19' N, 16$^\circ$ 05'
E), close to Capo Passero (South-East of Sicily), was identified
as an optimal candidate. The site is a wide abyssal plateau with an
average depth of about 3500 m, located at less than 80 km from the
shore and about 40 km from the shelf break (see Fig.
\ref{fig:capo_passero}).

\begin{figure}[h]
\begin{center}
\includegraphics[width=4.5 cm]{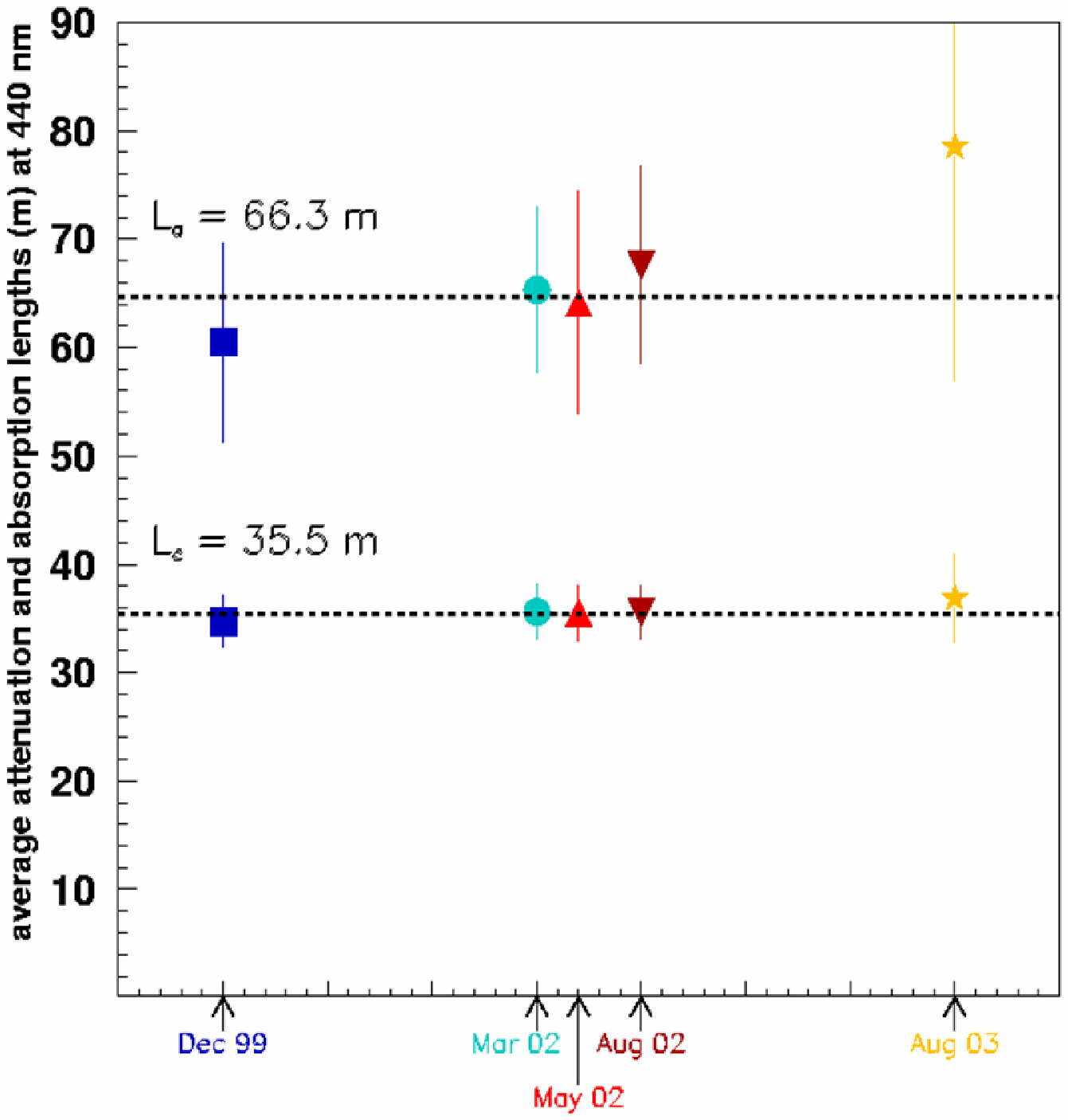}\hspace{1 cm}\includegraphics[width=7.2 cm]{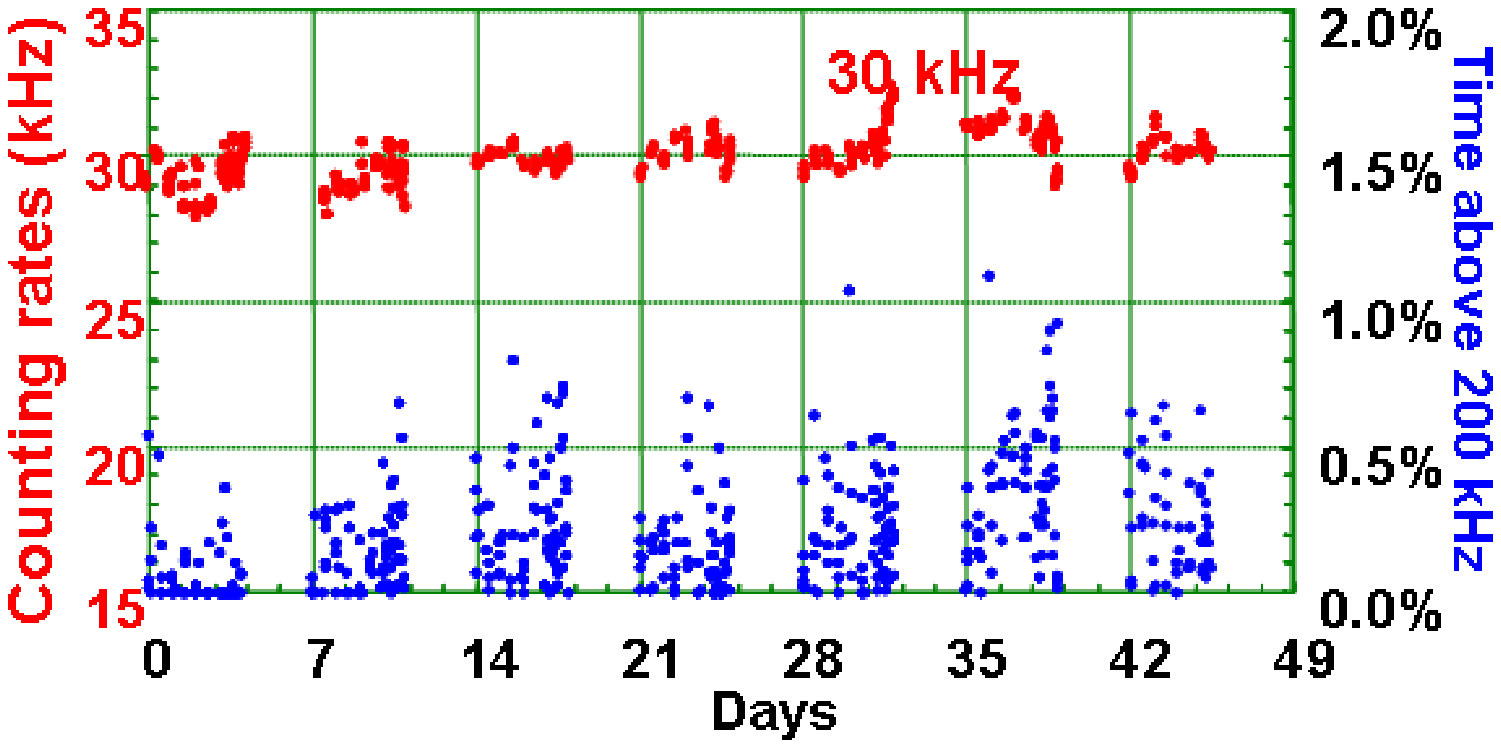}
\end{center}
\caption{\label{fig:absorbtion_attenuation-bioluminescence} Left: Absorbtion and attenuation coefficients measured in the Capo Passero site in several sea campaigns. Right: Optical Background rate measured in the Capo Passero site.}
\end{figure}

Water transparency was measured in situ using a set-up based on a transmissometer that
allowed the measurement of light absorption and attenuation at nine
different wavelengths ranging from 412 to 715 nm. The values of
the light absorption length measured at depths of interest for the
detector installation (more than 2500 m) are close to the one of
optically pure sea water (about 70 m at $\lambda$ = 440 nm) (see Fig.
\ref{fig:absorbtion_attenuation-bioluminescence}, Left).
Seasonal variations are compatible with the
instrument experimental error\cite{Riccobene06}. Another
characteristic of deep sea water that can affect the detector
performance is the optical background. This comes
from two natural causes: the Cherenkov light produced by electrons originated
by the $^{40}$K decay (the most abundant beta radioactive isotope
dissolved in seawater) and the so called bioluminescence, i.e.
light produced by biological organisms like bacteria or fish.
The first effect shows up as a constant rate background noise on the optical modules,
while the second one, when present, may increase the value of
the baseline rate and induce large fluctuations upon the
baseline. The bioluminescence effect can therefore cause long
DAQ dead time. In Capo Passero an average rate of about 30 kHz of optical noise
(measured with 10'' PMTs at 0.5 single photoelectron threshold) was
measured at a depth of 3000 m in several sea campaigns (see Fig.
\ref{fig:absorbtion_attenuation-bioluminescence}, Right). This value
is compatible with what expected from pure $^{40}$K background, with
rare high rate spikes due to bioluminescence, in agreement with
the measured vertical distribution of bioluminescent organisms
(the measurement indicate a low concentration of luminescent organisms at depths
greater than 2500 m). The programme of site characterization also
includes long term measurements of water temperature and salinity,
deep sea currents, sedimentation rate and bio-fouling. All data
confirm the optimal characteristics of the site\cite{NEMOAppec2003}.

\section{Feasibility study for the km$^3$ telescope}

\begin{figure}[h]
\begin{center}
\includegraphics[width=5.7 cm]{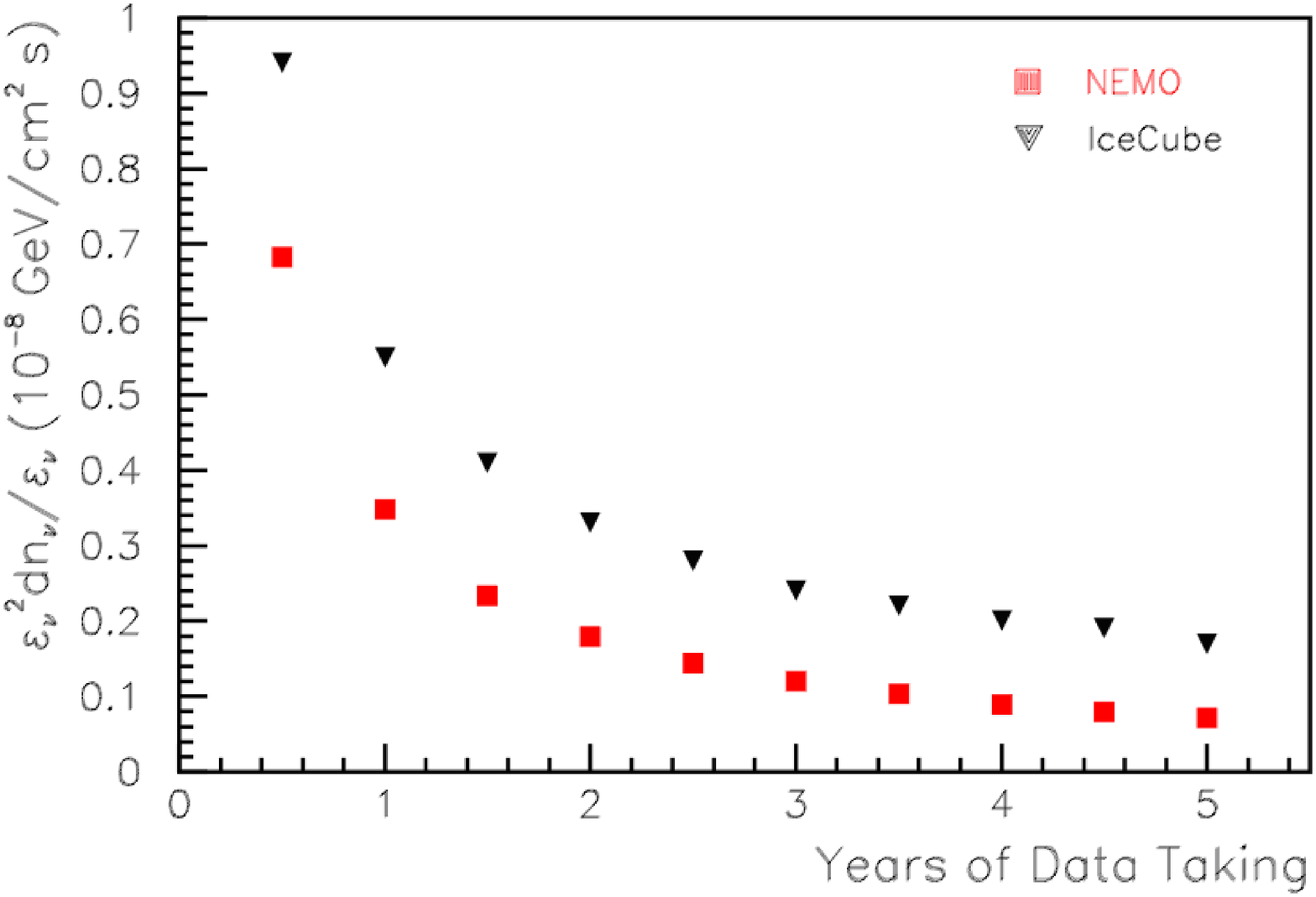}\hspace{1 cm}\includegraphics[width=5.7 cm]{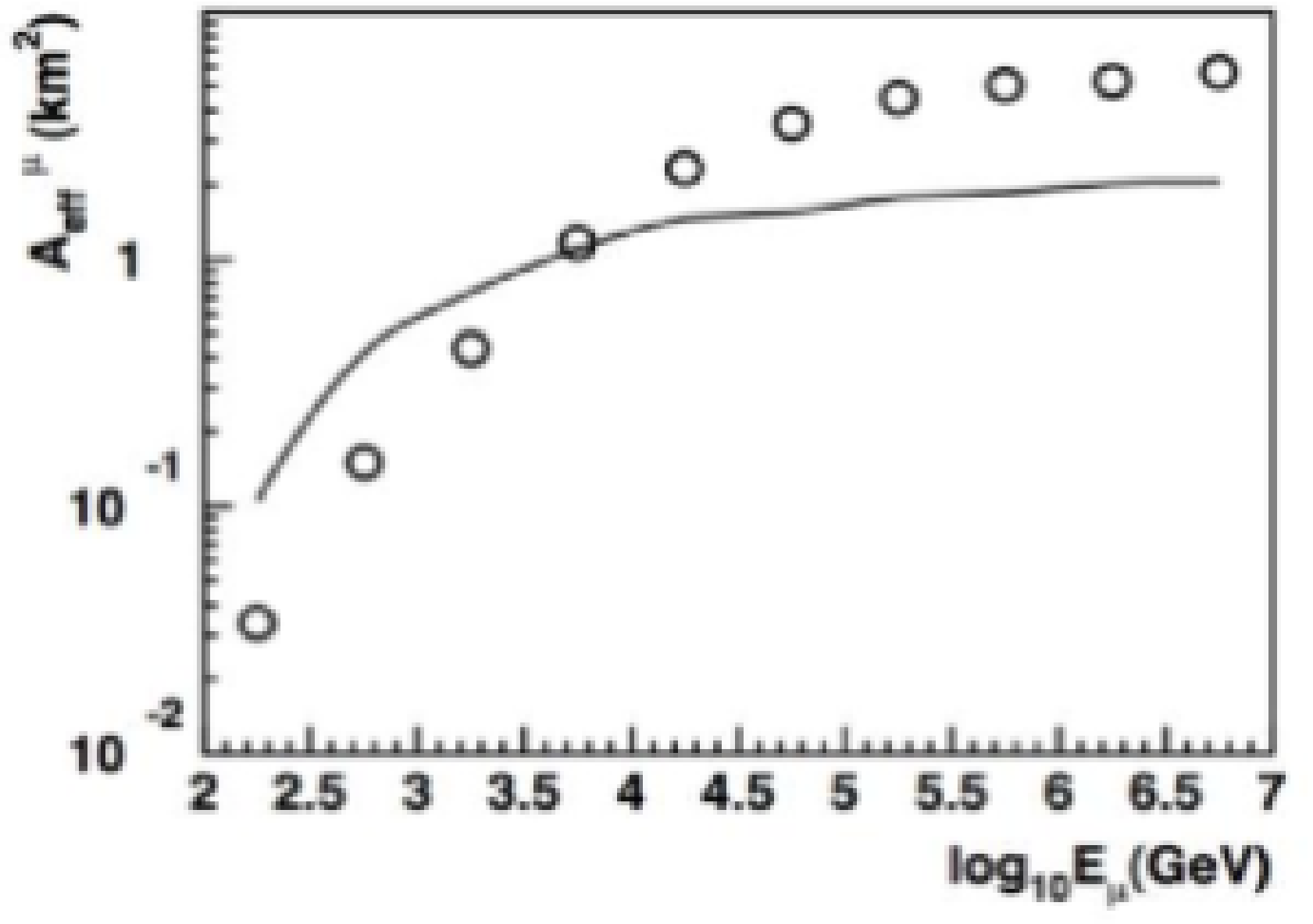}
\caption{\label{fig:sensitivity-effective_area}Left: Sensitivity to a neutrino spectrum with $\alpha=2$, coming from a $\delta=-60^\circ$ declination point-like source and
comparison with the IceCube detector. Right: Muon effective areas as a function of the muon energy for square lattices of NEMO towers 140 m spaced (continuous line) and 300 m spaced (open circles). }
\end{center}
\end{figure}

The design of an underwater km$^3$ neutrino telescope represents a
challenging task that has to match many requirements concerning the optimization of the physics discovery potentials, the detector performances, the technical feasibility and the project budget.
In general, a km$^3$ neutrino telescope is defined as an
array of about 5000 optical modules (OM) hosted on underwater
structures. NEMO proposes innovative structures: the NEMO {\it towers} designed to deploy, during a single operation, a large number of
OMs. Each floor of the tower hosts four optical modules
with 10" PMTs. Floors are arranged in a 3-dimensional structure in
order to locally allow event trigger and track reconstruction.
The proposed detector geometry consists of a squared array of
$9\times 9$ towers made of 16 floors each and with 5184 OMs.
The distance between the towers is 140 m.
The proposed architecture is "modular" and the layout can be reconfigured
to match different detector specifications. The detector performances were
evaluated by means of numerical simulations, using the software developed
by the ANTARES collaboration\cite{antares_codes} and adapted to a km$^3$ scale detectors\cite{Sapienza2003}. Using the above geometry and the site parameters measured in Capo Passero, Monte Carlo simulations show that the detector can reach an effective area $>$1 km$^2$ at muon energies of
about 10 TeV (see Fig. \ref{fig:sensitivity-effective_area}, Right), with an angular resolution of the order of few tenths of a degree. The effective area can be enhanced by increasing the distance between structures. In Figure \ref{fig:sensitivity-effective_area} the effective areas as a function of the muon energy are reported for two  $9\times 9$ detector configurations, with the same number of structures and OMs but different inter-tower distances of 140 m and 300 m. A significant gain is observed for the sparser array at the expenses of a higher energy detection threshold\cite{Coniglione06}. The sensitivity of the detector for a generic point-like source (see Fig. \ref{fig:sensitivity-effective_area}, Left)  is also reported as a function of the integrated data taking time. Simulations show that the expected sensitivity  to a point like astrophysical source at declination =-60$^\circ$ is about 1.2$\times$$10^{-9}$ GeV cm$^{-2}$ s$^{-1}$ (the used spectrum is $E_{\nu_{\mu}}^{-2}$), obtained for a search bin
of 0.3 degrees\cite{Distefano06}. The simulations indicate that
such a detector may reach a better sensitivity and smaller search
bin than IceCube\cite{ICECUBE2003}.

\section{NEMO Phase 1}

As an intermediate step towards the underwater km$^3$ detector and in order to validate the characteristics of the mechanical structure, of the
data transmission and of the power distribution system, we have developed a technological demonstrator called NEMO Phase 1, that includes prototypes of the critical elements of the proposed km$^3$ detector: the {\it junction box} and the {\it tower}. The prototype has been installed on December 2006 at
the Test Site of the Laboratori Nazionali del Sud at the depth of 2000 m 25 km offshore Catania.

\subsection{The Shore Station and the Underwater Station}

\begin{figure}[h]
\begin{center}
\includegraphics[width=11 cm]{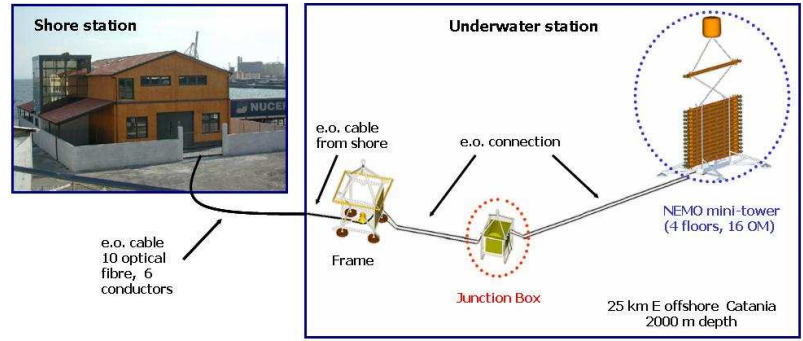}
\end{center}
\caption{\label{fig:sketch} A sketch of the NEMO Phase 1 Test Site.}
\end{figure}

The Test-Site infrastructure (see Fig. \ref{fig:sketch}) consists of 25 km of electro-optical submarine cable that connects the underwater station at 2000 m depth to the shore and a shore laboratory, located inside the Port of Catania. The shore station hosts the power feeding system, the instrumentation control system, the land station of the data transmission system and the data acquisition. The underwater cable is terminated in a frame hosting two e.o. connectors; two jumper-cables connect the frame to the  Junction Box and the Junction Box to a small scale tower, made of 4 floors, called {\it mini-tower}.

\begin{figure}[h]
\begin{center}
\includegraphics[width=6.5 cm]{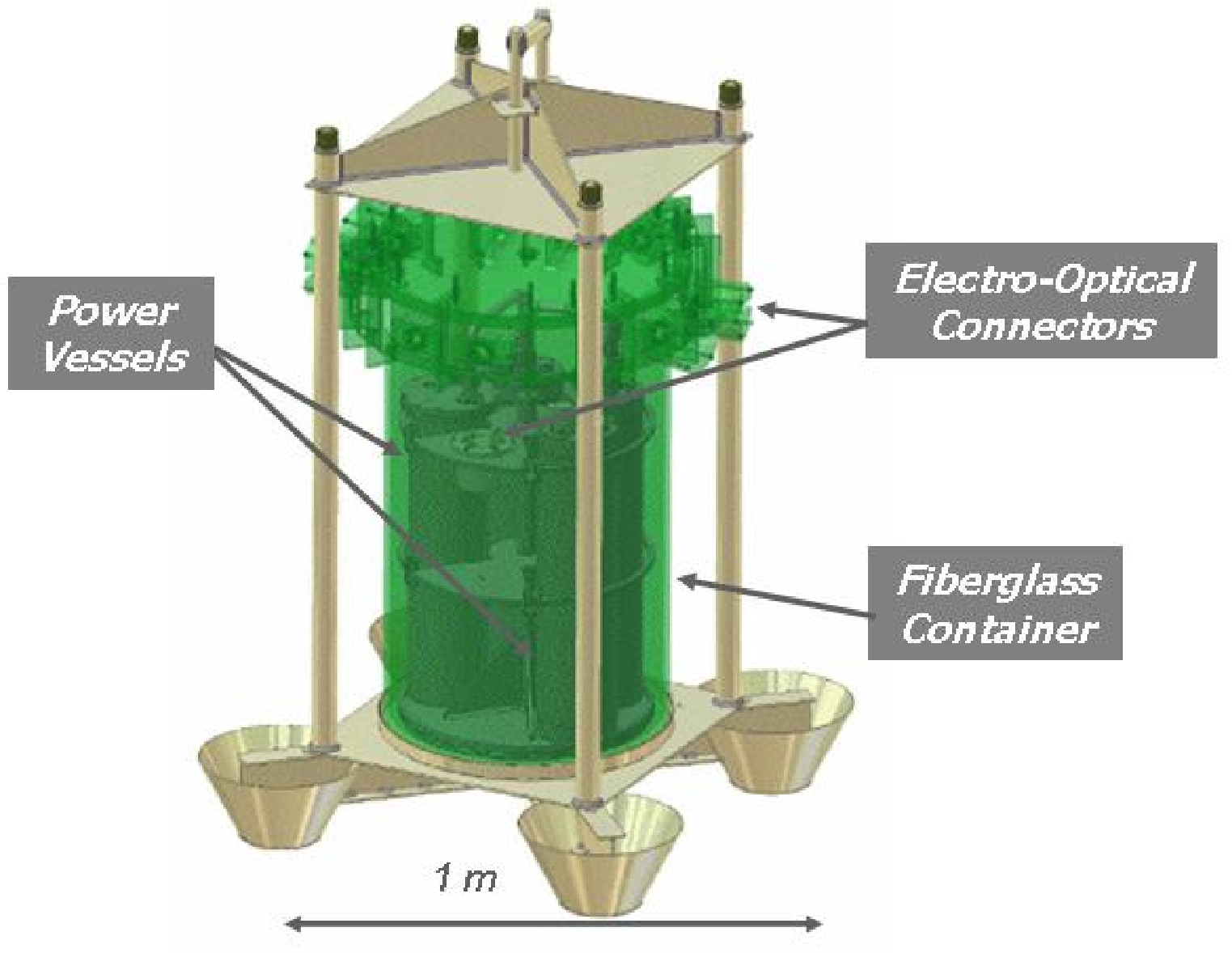}\hspace{1 cm}\includegraphics[width=3.8 cm]{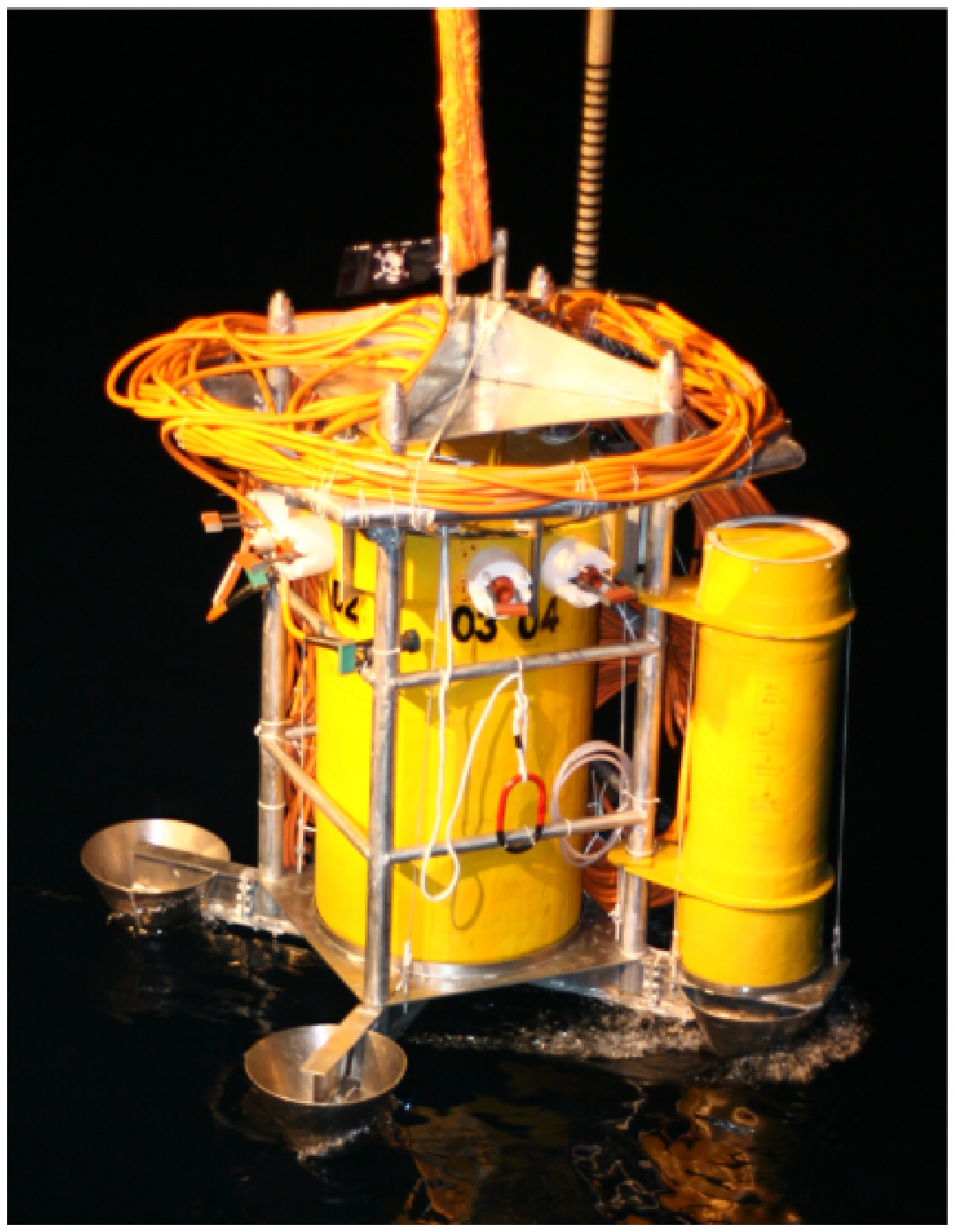}
\end{center}
\caption{\label{fig:JB} Left: A scheme of the Junction Box. Right: The Junction Box
during the deployment.}
\end{figure}

The Junction Box (JB) (see Fig. \ref{fig:JB})  is a key element of the detector. It provides connection between the main electro-optical cable and the instruments in the tower. It is designed to host and
protect, from the effects of corrosion and pressure, the opto-electronic boards dedicated to the distribution and control of the power supply and digitized signals.  The JB is realized with four cylindrical steel vessels hosted in a large fibreglass container to avoid direct contact between steel and seawater. The fibreglass container is filled with silicone oil and pressure compensated. All the electronics able to withstand high pressure is installed in oil bath, while the rest is located inside one pressure resistant container.

\begin{figure}[h]
\begin{center}
\includegraphics[width=6 cm]{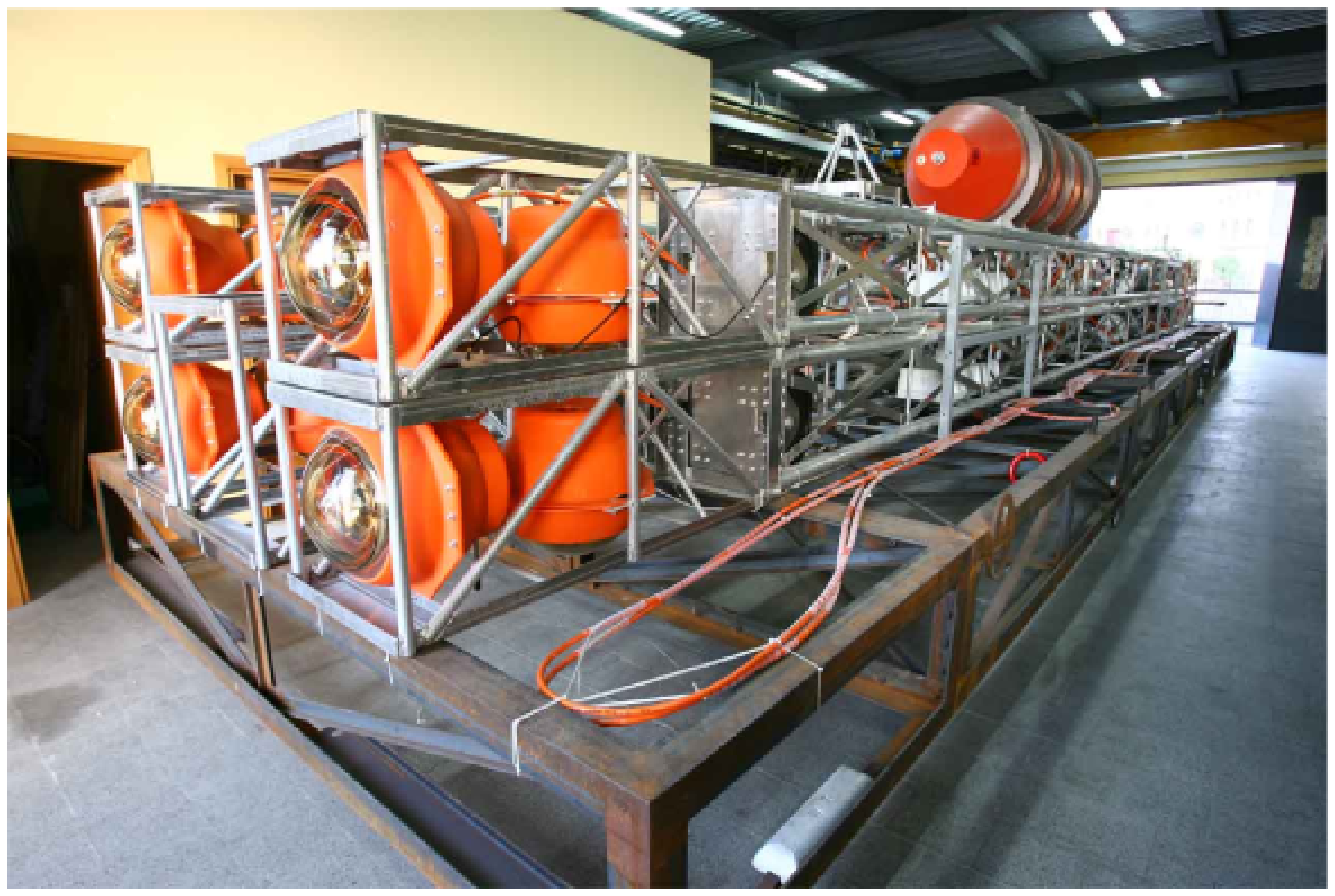}\hspace{1 cm}\includegraphics[width=2.5 cm]{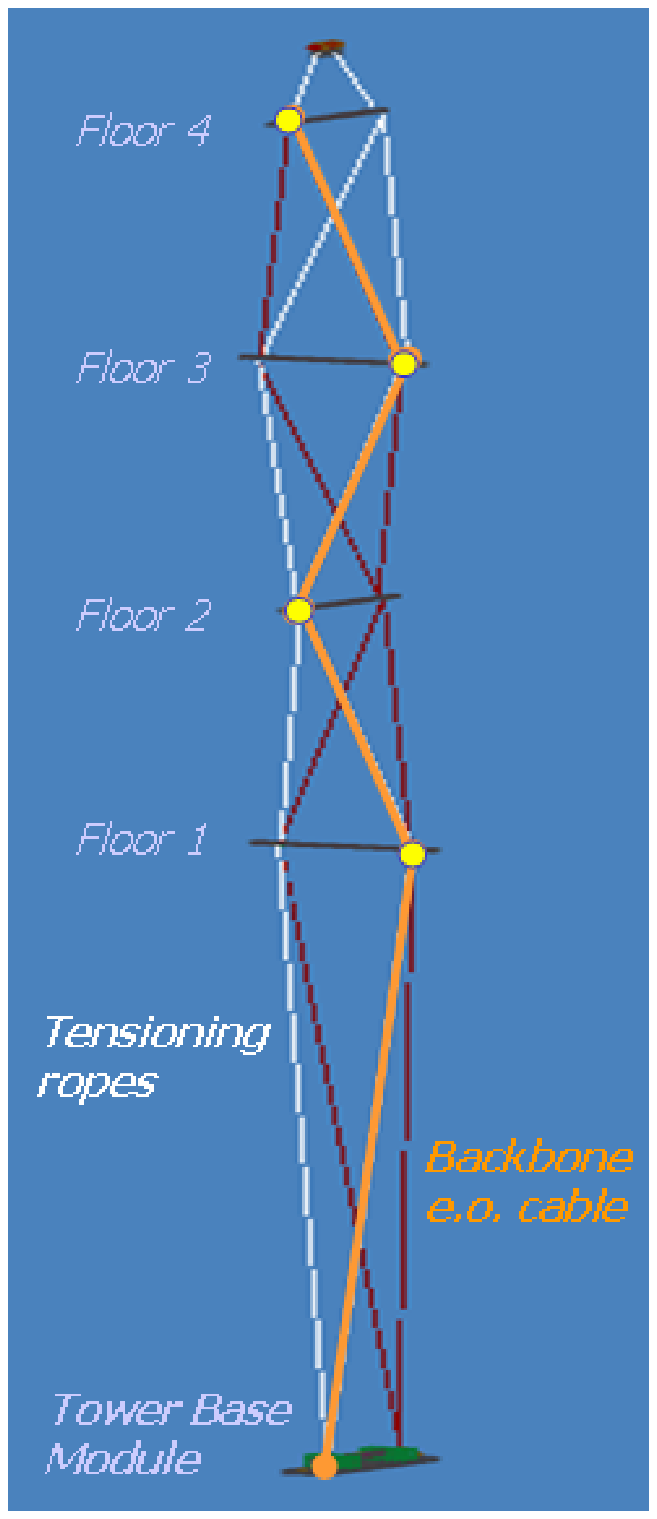}
\end{center}
\caption{\label{fig:minitower} Left: The mini-tower completely integrated at the Port of Catania. Right: A scheme of the mini-tower unfurled. }
\end{figure}

The tower that hosts the optical modules and the instrumentation is a three dimensional flexible structure composed by a sequence of floors interlinked by a system of cables and anchored on the seabed\cite{Musumeci}. The structure is kept vertical by appropriate buoyancy on the top. Each floor has a rigid aluminium structure, 15
m long, that hosts 4 optical modules (two down-looking and two horizontally-looking), environmental and control sensors and the front-end electronics. The vertical distance between floors is 40 m. An additional spacing of 100 m is added at the base of the tower, between the tower base and the lowermost floor to allow for a sufficient water volume below the detector. One of the advantages of this structure is the fact that it can be compacted, by piling each
storey upon the other, to allow easier transport and deployment. After
deployment the structure is unfurled (see Fig. \ref{fig:minitower}, Right): each floor will be rotated by 90$^\circ$, with respect to the up and down adjacent ones, around the vertical axis of the tower. Each floor is connected to the following one by means of four ropes that are fastened in a way that forces each floor to take an orientation perpendicular with respect to the adjacent (top and bottom) ones.
While the design of a complete tower for the km$^3$ foresees 16 floors, the prototype realized for the Phase-1 project is a mini-tower of 4 floors (see Fig. \ref{fig:minitower}, Left).
The instrumentation installed on the mini-tower includes several sensors for calibration and environmental monitoring. In particular two hydrophones are mounted on the tower base and at the extremities of each floor. These, together with an acoustic beacon placed on the tower base and other beacons installed on the sea bed, are used for precise determination of the tower position by means of time delay measurements of acoustic signals. The other environmental probes are: a Conductivity-Temperature-Depth (CTD) probe used for the monitoring of the water temperature and salinity, a light attenuation probe (C*) and an Acoustic Doppler Current Profile (ADCP) that will provide continuous monitoring of the deep sea currents along the whole tower height.

Inside each floor structure are installed a Power Floor Control Module (PFCM) and a Floor Control Module (FCM). The latter is the core of the system since it hosts all the floor electronics for data transmission. The PFCM turns the high voltage tension of 380 VAC in continuous tension and furnished the three tension of 4 V, 6 V and 12 V (DC) to the front-end electronics and to the oceanographical probes. It is hosted in a silicone oil filled plastic container, since all the power supply subsystem has been tested to operate under pressure\cite{Cocimano}. The FCM is connected to
the OMs by means of four electro-optical cables, and to the floor instrumentation (oceanographic probes, hydrophones for the acoustic positioning system) via electrical cables.
Each optical module is composed by a 10" photo-multiplier enclosed in a 17" pressure resistant sphere of thick glass. The used PMT is an 10" Hamamatsu R7081Sel with 10 stages. In spite of its large photocathode, the used PMT has a time
resolution of $\sim$3 ns FWHM for single photoelectron pulses
and a charge resolution of $\sim$$35\%$. Mechanical and optical contact between the PMT and the internal glass surface is ensured by an optical silicone gel. A $\mu$-metal cage shields the PMT from the Earth's magnetic field. The base card circuit for the high voltage distribution (Iseg PHQ 7081SEL) requires only a low voltage supply (+5 V) and generates all necessary voltages for cathode, grid and dynodes with a power consumption of less than 150 mW.
A front-end electronics board is also placed inside the OM. Sampling at 200 MHz is accomplished by
two 100 MHz staggered Flash ADCs, whose outputs go to an FPGA which classifies (according to a remotely programmable threshold) the signal as valid or not, stores it with an event time stamp in an internal 12 kbit FIFO, packs OM data and local slow control information, and codes everything into a bit stream frame ready to be transmitted to the FCM at 20 Mbit/s rate. The FCM on each floor collects data from the floor OMs and the auxiliary instrumentation, creates a proprietary optical link for data transmission at about 155 Mbps toward the shore laboratory. From the opposite direction, the FCM receives also slow control data, commands and auxiliary information, and the clock and synchronizations signals needed for apparatus timing.

\subsection{First results}

The data stream reaches shore and it is managed by the shore DAQ systems. A Run-Console acquires the "raw" data form the OMs and sends them to a Master-CPU, which performs the triggering. We have implemented four trigger types: 1) simple coincidence with $\Delta$t=20 ns; 2) floor coincidence, between two PMTs of the same floor placed into different side, with $\Delta$t=200 ns; 3) charge threshold (3.5 p.e.); 4) random trigger ($\Delta$t=1 ms).
Post-Trigger data are sent to a data storage and recorded into a database (the Data Manager). The database also records the detector parameters, the acoustic positioning data and the environmental sensors data. A Run Control program continuously polls the database allowing real-time monitoring of PMT rates,  environmental sensors data and all detector control parameters.

\begin{figure}[h]
\begin{center}
\includegraphics[width=6.5 cm]{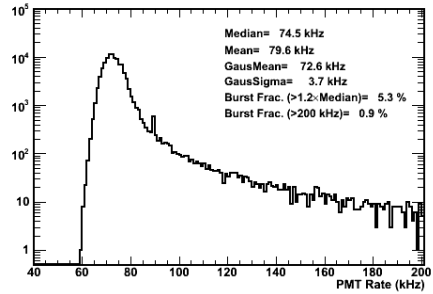}
\end{center}
\caption{\label{fig:results} Average optical background rate measured by PMTs installed at the Nemo Phase 1 Test Site.}
\end{figure}

\begin{figure}[h]
\begin{center}
\includegraphics[width=7.5 cm]{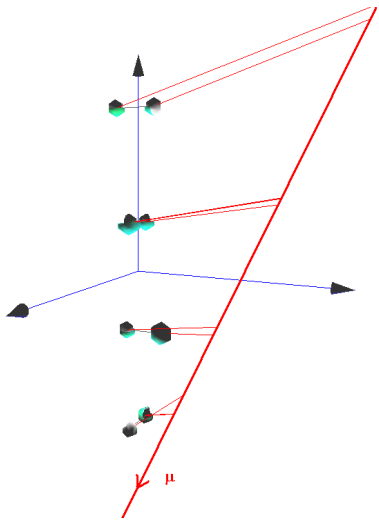}
\end{center}
\caption{\label{fig:track} One of the first atmospheric muon tracks reconstructed.}
\end{figure}

The average measured rates (see Fig. \ref{fig:results}) are about 75 kHz for all PMTs as expected from previous optical background measurement performed in the site.
Simulations and track reconstruction programs are presently running to analyse data. First atmospheric muon tracks (see Fig. \ref{fig:track}) have been reconstructed.

\section{NEMO Phase 2}

Although the Phase-1 project has provided a fundamental test of the technologies proposed for the realization and installation of the detector, these results must be finally validated at the installation depths of the km$^3$ detector $>$3000 m.
For these motivations the realization of an infrastructure at the Capo Passero site (NEMO Phase 2) has been undertaken. It will consist of a 100 km cable, linking the 3500 m deep sea site to the shore, a shore station, located inside the harbour area of Portopalo di Capo Passero, and the underwater infrastructures needed to connect prototypes of the km3 detector.
Due to the longer cable needed, a different solution, with respect to the Phase-1 project, for the electro-optical cable was chosen. In this case the backbone cable (manufactured by Alcatel) will carry a single electrical conductor, that can be operated at 10 kV DC allowing a power transport of more than 50 kW, and 20 single mode optical fibres for data transmission\cite{Sedita}.
The completion of this project is foreseen by the end of 2008. At that time it will be possible to connect one or more prototypes of detector structures, allowing a full test at 3500 m of the deployment and connection procedures. Phase 2 project will also allow a continuous long term on-line monitoring of the site properties (light transparency, optical background, water currents, ...) whose knowledge is essential for the installation of the full km3 detector.

\section{Conclusions}

The design of the Mediterranean km$^3$ telescope for high
energy astrophysical neutrinos is a challenging task; several
collaborations in Europe are working on the realisation of demonstrators. More efforts are needed to develop a project for the km$^3$ detector; for this reason EU funded KM3NeT, which is expected to provide a design
study for the Mediterranean km$^3$ telescope within 2009\cite{km3net}. The activities of
the NEMO collaboration are contributing to this goal: an extensive
study on the site properties has
demonstrated that the Capo Passero site has optimal characteristics for
the telescope installation; a technical feasibility analysis for
the km$^3$ detector has shown that a detector with effective area
for TeV muons of $\sim$1 km$^2$ is realisable at an affordable cost. A technological demonstrator has been installed at the underwater Test Site of the LNS in Catania on December 2006 and is now taking data. First muon tracks were already reconstructed.

\end{document}